# Buffering blood pressure fluctuations by respiratory sinus arrhythmia may in fact enhance them: a theoretical analysis

T.Buchner, J.J.Żebrowski

Faculty of Physics, Warsaw University of Technology ul. Koszykowa 75, 00-662 Warsaw, Poland

email: buchner@if.pw.edu.pl

G.Gielerak

Departament of Internal Diseases and Cardiology, Military Institute of the Health Services, Szaserów 128, 00-909 Warsaw, Poland

Abstract

Using a three-compartment model of blood pressure dynamics, we analyze theoretically the short term cardiovascular variability: how the respiratory-related blood pressure fluctuations are buffered by appropriate heart rate changes: i.e. the respiratory sinus arrhythmia. The buffering is shown to be crucially dependent on the time delay between the stimulus (such as e.g. the inspiration onset) and the application of the control (the moment in time when the efferent response is delivered to the heart). This theoretical analysis shows that the buffering mechanism is effective only in the upright position of the body. It explains a paradoxical effect of enhancement of the blood pressure fluctuations by an ineffective control. Such a phenomenon was observed experimentally. Using the basis of the model, we discuss the blood pressure variability and heart rate variability under such clinical conditions as the states of expressed adrenergic drive and the tilt-test during the parasympathetic blockade or fixed rate atrial pacing. From the results of the variability analysis we draw a conclusion that the control of blood pressure in the HF band does not directly obtain the arterial baroreceptor input. We also discuss methodological issues of baroreflex sensitivity and sympathovagal balance assessment.

## INTRODUCTION

The relation between the heart rate (HR) and the arterial blood pressure (ABP) is a subject of intensive experimental and theoretical study (see the review [1]). We introduce a simple but nonlinear compartmental model of short-term cardiovascular dynamics in order to explain a rarely discussed experimental observation. Elstad et al. [2] found that the respiratory sinus arrhythmia (RSA) has an opposite effect on the systolic and on the mean arterial pressure in the supine position but not in upright humans. Another intriguing observation is that the parasympathetic blockade [3] as well as the fixed rate atrial pacing [4] reduces ABP oscillations in supine but not in the upright position. The model we present shows qualitatively similar effects and provides a good starting point for further discussion.

Respiratory sinus arrhythmia is primarily mediated by the respiratory-related, efferent parasympathetic activity [5]. This activity is often called "vagal tone", which expresses its tonic (i.e. constant) character. It seems, however, that this vagal tone has an important variable component related to the phase of respiration [6]. This variable component acts periodically on the sinus node. This action is close to periodic and has two effects: the mean HR is reduced and a fluctuation of the instantaneous HR around its mean value occurs (respiratory sinus arrhythmia). Both effects may be temporarily eliminated by cooling the vagus nerve [7]. Experimental studies suggest that the reduction of HR may be caused by an entrainment of the subordinate heart rhythm to the external (respiratory) rhythm [5]. Another dynamical effect was observed experimentally [5] as well as reproduced in mathematical models [8,9]. It is the so-called "vagal paradox": that the increasing frequency of the parasympathetic drive in some ranges of respiratory frequency causes a paradoxical acceleration (instead of the generally observed deceleration) of HR. On the other hand, under certain conditions, the respiration-related variability is also present in muscle sympathetic nerve activity (MSNA) recordings [10] and the role of baroreflex in RSA is considered to be important [11]. Last but not least, RSA was also observed in denervated hearts indicating a contribution from mechanical coupling [12]. Thus, although RSA is a well known phenomenon, there seem to be a number of distinct mechanisms responsible for the occurrence of HR variability in the respiratory frequency band (high frequency – HF). Below, the effect of the heart rate on breathing will be denoted for brevity "vagal feedback" but we do not try to conclude on its actual nature.

The respiratory-related fluctuations of selected cardiovascular quantities were directly measured using Doppler technique [13,14,2]. Paper [13] describes the respiratory-related fluctuations of arterial blood pressure ABP. They seem to be caused by the transmural mechanical action of the intrathoracic pressure on large vessels that modulates the stroke volume (SV) [14]. The second paper [2] reported that the mechanism of buffering of respiratory-related ABP fluctuations by the heart rate exhibits unexpected properties. Not only was the buffering dependant on the body position, but also it seemed to act differently on different properties of the ABP. Only the mean arterial pressure (MAP) fluctuations were buffered, while the systolic arterial pressure (SAP) fluctuations were actually enhanced in the supine position of the body. Similar result was reported in [34]. As the enhancement of respiratory-related fluctuations of the ABP does not seem to have an apparent physiological role, we rather interpret this effect as an intriguing inefficiency of buffering in the supine position.

In this paper, we discuss a possible mechanism underlying the phenomenon of selective buffering and its dependence on the position of the body. The mechanism is studied in a simple model of cardiovascular dynamics consisting of three compartments with nonlinear elements. The motivation for considering such a minimalistic model comes from the classical paper on the mean circulatory filling pressure [15] where the whole cardiovascular system was reduced to only two segments.

## MODEL DEVELOPMENT

The model, defined by Eq. 1-10 consists of only three compartments and the glossary of parameters used in the model with their nominal values is summarized in Table 1.

Table 1

| Symbol | Quantity | Nominal value |
|---|---|---|
| $C_A$ | Systemic arteries compliance | 1.6 ml/mmHg |
| $C_V$ | Systemic veins compliance | 100 ml/mmHg |
| $C_C$ | Effective cardiopulmonary compliance | 4.3 ml/mmHg |
| $Z_{VC}$ | Ventricular outflow conductance | 200 ml/(s mmHg) |
| $Z_{AV}$ | Arterial outflow conductance (1/TPR) | 1.1 ml/(s mmHg) |
| $Z_{CA}$ | Effective cardiopulmonary outflow conductance | 166 ml/(s mmHg) |
| $p_{I0}$ | Intrathoracic pressure | -4 mmHg |
| $p_{I1}$ | Intrathoracic pressure variability | 2 mmHg |
| $r_{HR0}$ | Resting heart rate | 66 bpm |
| $r_0$ | vagal control amplitude | 1s |
| $\tau$ | vagal control delay | 0.3 s |
| $\tau_v$ | vagal control decay rate | 0.3 s |
| $r_{BR0}$ | Resting respiratory rate | 20 bpm |

We consider an arterial compartment (AC), a cardiopulmonary compartment (CC) and a venous compartment (VC) with the respective blood pressures denoted by $p_a$, eq. (1), $p_c$, eq. (2) and $p_v$, eq. (3):

$$C_a \frac{dp_a}{dt} = \sum_i \delta(t - t_i) Z_{ca}(p_c - p_t) - Z_{av} p_{av} \quad (1)$$

$$C_c \frac{dp_c}{dt} = \max(Z_{vc} p_{vc}, 0) - \sum_i \delta(t - t_i) Z_{ca}(p_c - p_I) \quad (2)$$

$$C_v \frac{dp_v}{dt} = Z_{av} p_{av} - \max(Z_{vc} p_{vc}, 0) \quad (3)$$

The pressures with double indices represent blood pressure gradients between neighboring compartments (e.g. $p_{av}=p_a-p_v$). Each compartment is characterized by its compliance and outflow conductance (the reciprocal of resistance). $Z_{xy}$ denotes the conductance between the compartments $x$ and $y$, whereas $p_I$ stands for intrathoracic pressure. The inertial effects due to blood flow are neglected, following [16]. The normative values of parameters listed in Table 1, apart from those that characterize the vagal control, were taken from [16]. The properties of the cardiopulmonary compartment were set arbitrarily in such a way as to obtain a reasonable range of blood pressure changes in the arterial and the venous compartment. The model was implemented in Matlab, extending the RCVSIM framework [17] and separately in C#. Both implementations gave quantitatively indistinguishable results. The structural stability of the model was verified by varying the parameters by ±10% about their nominal values. The HR and the ABP for all parameter values were within the physiological limits.

The mathematical form of (1) and (2) resembles the "kicked" oscillator models which where extensively studied in nonlinear dynamics [18] and may be converted to discrete time systems, such as described in ref. [19] or [20].

*The cardiopulmonary compartment*

With respect to the model of the cardiovascular system defined in [16] and other models referenced therein, as well as the model discussed in [15], the introduction in the proposed model of a surrogate cardiopulmonary compartment (2) is new. This compartment represents the function and properties of the heart, the valves and the large vessels situated in the thorax. The compartment is a surrogate that allows to treat the dynamics within the pulmonary circulation and the heart as a single entity.

As the cardiopulmonary compartment is anatomically located in the thorax, the external (transmural) pressure which acts on the vessels located within this compartment is the intrathoracic pressure $p_I$ (cf (2)). It sets the reference pressure that affects the outflow from this compartment. Such approach is widely used by other researchers [16,21,22] who consider elastic properties of the vascular wall. The intrathoracic pressure $p_I$ changes with the phase of breathing, as defined by equation (4):

$$p_I(t) = p_{I0} + p_{I1}(1 + \cos 2\pi \Phi) \qquad (4)$$

The inflow into the cardiopulmonary compartment is represented in (2) by a conductance $Z_{vc}$, which may be altered in order to model orthostatic changes in the venous return. Following [16], the action of valves is mimicked by an ideal diode (the function max in (2) and (3)) to force a single direction of the flow.

*Modeling the hemodynamic effect of a heart action*

The heart rate is modeled using an "integrate and fire" type of model. Extending the idea of [23] and [24], we introduce a single variable $\varphi(t)$ with the range $[0,\infty[$, that represents the phase of the cardiac cycle.

$$\frac{d\phi}{dt} = r_{HR0} + \sum_j f(t - t_j - \tau) \cdot F(\phi \bmod 1) \qquad (5)$$

Functions $f$ and $F$ in (5) represent the neural control of the heart rate and the sum is over all breaths, as explained in the next paragraph. For convenience the phase $\phi$ is normalized to 1 instead of $2\pi$.

The heartbeats are generated in the moments in time $t_i$ when the phase exceeds any integer value:

$$t_i : \phi(t_i) = i \qquad (6)$$

The hemodynamic effect of heart action is a short, instantaneous flow pulse in eq. (1),(2). We found such a flow to be qualitatively similar to that obtained in the model of varying elastance [16]. (Note that the derivative of the varying elastance has a shape which may be approximated, with a reasonable accuracy, by the Dirac delta function. – c.f. [16])

The Sterling law is not directly introduced into the model. The effect of the cardiac filling time is, however, taken into account in the model because the BP in the cardiopulmonary compartment $p_c$ changes approximately reciprocally to the BP in the arterial compartment $p_a$. If the RR interval is short and the $p_a$ at the end of diastole is still high, the BP in the cardiopulmonary compartment $p_c$ will be low which marks an incomplete filling and, in consequence, the next stroke volume (second term on the r.h.s. of eq.(2)) will be reduced. Such an approximation seems sufficient as long as the amplitude of the heart rate fluctuations is not large.

The stroke volume in our model does not depend directly on the arterial pressure. The only dependence is indirect: through the cardiac filling time, as explained above. The value of $p_l$ is modulated by breathing in eq. (4), which results in a direct modulation of the cardiopulmonary outflow (stroke volume) and hence in a modulation of the pulse pressure (PP) and SAP (1) with the frequency of respiration (HF, i.e. 0.3 Hz). In our model this modulation is due to purely mechanical effects.

*The respiratory rhythm and neural control of the heart rate*

The respiratory rhythm is introduced as another "integrate and fire" oscillator, with the dynamics of its phase variable $\Phi(t)$ described by equation (7):

$$\frac{d\Phi}{dt} = r_{BR0} \qquad (7)$$

The respiratory rate is taken to be constant. The firing times for breathing are denoted by $t_j$.

$$t_j : \Phi(t_j) = j \qquad (8)$$

The main neural feedback considered in the model is in the form of the respiratory-related vagal activity

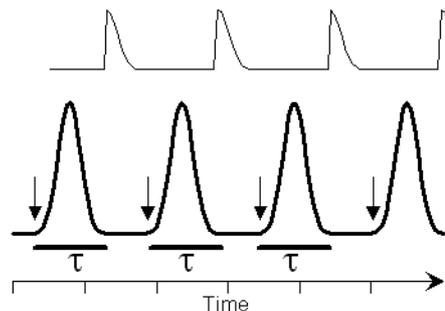

Fig. 1. Schematic of the vagal input in the kicked Windkessel model. The top curve is the simulated vagal control signal (thin line), the bottom curve is the simulated respiratory signal. (thick line). The spikes of vagal control are delivered at delay time τ (marked by the thick horizontal lines) after inspiration onsets, marked by arrows.

(second term in eq. (5)) that synchronizes the heart phase with the breathing phase. We refer to it as vagal but as discussed in the preceding chapter, it may model any type of respiratory-related activity that acts on the heart. The role of this activity is to control the respiratory-related HF variability of ABP by adjusting the instantaneous heart rate. The effect of vagal activity on the cardiac pacemaker was shown to be phase-dependant [5]. Following ref. [24], we use the phase effectiveness function in the form defined by eq. (9).

$$F(\phi) = \phi^{1.3}(\phi - 0.45) + \frac{(1-\phi)^3}{(1-0.8)^3 + (1-\phi)^3} \qquad (9)$$

The vagal input in this model is simplified: it is added to the cardiac phase as a spike, with a subsequent exponential decay, at time $\tau$ after each inspiration onset (Fig. 1), eq. (5). The cardiac oscillator responds by a lengthening or a paradoxical shortening ("vagal paradox") of the cardiac cycle.

The mathematical form of such a decaying spike is described as the function $f$ in eq. (10),

$$f(t) = r_0 \cdot \Theta(t) \cdot \exp(-t/\tau_v) \qquad (10)$$

which consists of the exponential decay multiplied by the Heaviside theta function to ensure that the activity is nonzero only for time positive (c.f. 5). The neural control of the heart rate has two parameters: the amplitude of the spike $r_0$ and the delay time $\tau$ (with respect to the inspiration onset) at which it is delivered.

RESULTS

The numerical output of the model is shown in Fig. 2. The waveform of $p_a$ follows the Windkessel principle: i.e. after each flow pulse ("kick") there is an outflow to the venous compartment, proportional to the pressure gradient $p_{av}$.

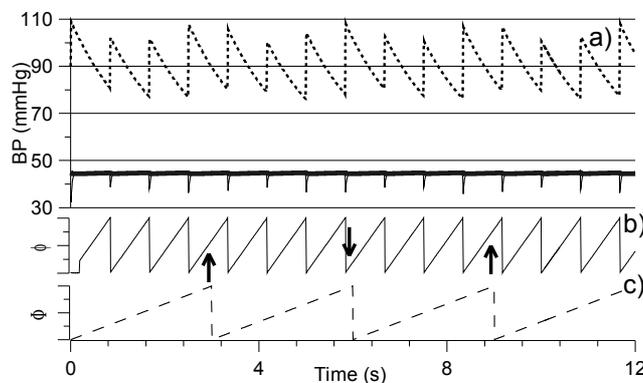

Fig. 2. Example of numerical output from the model. Part a) shows blood pressure for all compartments: $p_a$ of the arterial compartment AC (dashed curve), $p_v$ of the ventricular compartment VC (thick curve), and $p_v$ of the cardiopulmonary compartment CC (downward spikes marked by a thin curve overlapping $p_v$). Blood pressure in the VC is constant in this scale of the plot. Part b) shows the phase of the heart. Part c) shows the phase of the respiratory rhythm (dashed curve) reduced to the [0,1] interval. Arrows mark moments of the application of control..

The magnitude of $p_a$ is within physiological limits. The $p_a$ wave envelope modulation (see dashed curve in fig.2a) is due to the respiratory-related changes of the pulse pressure.

The pressure in the venous compartment $p_v$ is nearly constant because of the venous compliance $C_v$ which is two orders of magnitude larger than the compliances of the other two compartments. Its value is above the physiological level which seems justified given the qualitative character of the model. Blood pressure in the cardiopulmonary compartment $p_c$ exhibits sudden decreases (downward spikes marked by a thin line overlapped by the thick curve in fig.2a) caused by the instantaneous outflow to the arterial compartment at each heartbeat. During the diastole blood pressure slowly increases due to the inflow from the venous compartment. Although the magnitude of $p_c$ obtained is unrealistic, it retains such physiological property that the stroke volume depends on the time of the preceding diastole.

The bottom traces (fig.2b and fig.2c) show the respiratory phase and the cardiac phase, respectively. Note the small changes of the cardiac phase (arrows in fig.2b) after each respiration onset, almost invisible in this scale of the plot. At these points, the neural control of the heart rate (second term on the r.h.s. in (5)) is applied.

*Effectiveness of buffering*

In order to study the effectiveness of buffering, the parameter space of the model was explored. The two parameters that describe the neural control are the amplitude of the stimulus $r_0$ and the delay $\tau$. The effectiveness of the buffering as a function of each of these two parameters is shown in Fig. 3 and Fig. 4, respectively.

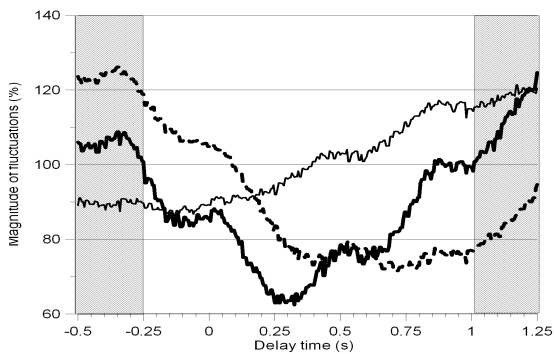

Fig. 3. Magnitude of the fluctuations of BP as function of the delay time $\tau$. Three BP variables are shown: MAP (thick solid curve), SAP (thin curve) and DAP (thick dashed curve). Grey area marks the ranges of an ineffective control of the MAP

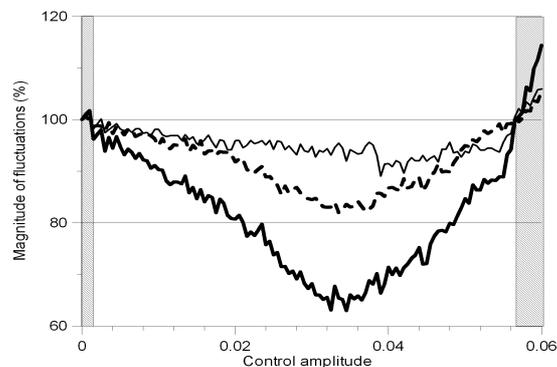

Fig. 4. Magnitude of the fluctuations of BP as a function of the control amplitude $r_0$. Three BP variables are shown: MAP (thick solid curve), SAP (thin curve) and DAP (thick dashed curve). Grey area marks ranges of an ineffective control of the MAP

The effectiveness of buffering is measured using the relative magnitude of the fluctuations calculated in the following way.

First, the standard deviation of each of the blood pressure signals is determined in the absence of the neural control but with the respiratory modulation present. In this way, the reference values of the standard deviation for the ABP-related variables: SAP, DAP and MAP are obtained. The MAP is not calculated from the full waveform. Instead, we use the widely used formula: MAP=(2 DAP + SAP)/3 which seems to provide a reasonable approximation. As there is only one source of fluctuations in the system and a single buffering mechanism, there is no need to separate the power of the fluctuations into different frequency bands. Therefore, the standard deviations, which reflect the total power of the fluctuations, seem to be a sufficient measure to assess the buffering effectiveness.

Next, for each value of the control parameter, the model is iterated for 20 seconds with the neural control turned on. From the model output, the standard deviation of the ABP-related variables is calculated, divided by their reference values and expressed in percent. This procedure gives a relative magnitude of ABP fluctuations. If the control is effective, this magnitude is smaller than 100%. If the control is ineffective and it paradoxically enhances the fluctuations, the magnitude exceeds 100%. We do not consider the absolute changes of the ABP-related variables due to the control mechanism as in real applications the depth of respiratory-related modulation depends on many factors such as tidal volume, respiratory frequency, BMI, stiffness of vasculature, the actual intrathoracic pressure, the presence of pathologies such as emphysema, which dramatically changes the response to the Valsalva manouvre [25] and many others.

In Fig. 3, it may be seen that the range of effective control for all the ABP-related variables is different. The one that seems to be most important, i.e. the MAP, is effectively buffered when $\tau$ is in the range [-0.25,1] s. It seems that in the middle of this range, approximately at $\tau_c$=0.3 s, the effective control range of SAP ends. Below this critical value, the control is effective for both SAP and MAP i.e. the fluctuations of both of them will be reduced. Above this threshold, only MAP is effectively controlled. DAP has a completely different range of effective control. It starts for delay times below $\tau_c$ and ends well above the right extreme value yielding an effective control for the SAP.

Fig. 4 depicts the expected result that for each controlled variable there exists a single minimum of the fluctuations with respect to the amplitude of the control $r_0$. For the amplitude of control that is smaller than a certain value, the HR fluctuations are too small to effectively buffer the ABP changes. For $r_0$ which exceeds the effective buffering range, the magnitude of the control is too large and introduces large fluctuations that are not necessarily in counterphase with those due to the mechanical modulation.

## DISCUSSION

*Selective control of the blood pressure*

The first result to discuss is the finding that the control seems to act differently at each of the ABP-related variables. There exists a range of delay times in which the control is effective only for MAP and not for SAP. A selective control of blood pressure in the HF range was experimentally observed in ref. [26] and ref. [3] in which the effect of a total cardiac autonomic blockade was discussed. It was observed that the blockade acts mainly on the fluctuations of diastolic BP, leaving the fluctuations in systolic BP quite unaltered. The popular approximate empirical formula for the mean arterial pressure: *MAP = (2 DAP + SAP)/3* is the result of observations and has a simple consequence: for the stabilization of MAP it is two times more important to stabilize the diastolic AP than the systolic. Therefore, it is possible, that within a certain parameter range, the fluctuations of the systolic AP will be enhanced and those of the mean AP – reduced – due to the action of the same control mechanism. The destabilization of SAP would then appear as a by-product of the control of fluctuations of MAP.

Another experimental result that may be quite reliably explained using current theory is the presence of positional changes in BP variability amplitude, discussed in ref. [2] if we assume that the delay time $\tau$ in (5) depends on the position of the body. This assumption seems to be supported by the fact that the RSA pattern and the relative phase between ABP and HR both depend on the position of the body [27,28,34]. With the change of the body position not only the RSA pattern would change, but also the whole pattern would be shifted by 1 s with respect to the inspiration onset. This is equivalent to the change of the delay time $\tau$.

The explanation of the exact physiological mechanism that causes the change of the RSA pattern observed in [27] is outside the scope of this work. One may only hypothesize that it originates from the change of the vertical distribution of the pulmonary blood flow in relation with the position of the body. Such a change could introduce a phase shift between the phase of breathing and the phase of modulation of blood pressure in the cardiorespiratory compartment. On the ground of our model we may consider this phase shift to be included in the delay time $\tau$.

In subsequent paragraphs, a number of issues related to the ABP control itself is discussed. The notion that HR actively buffers the ABP fluctuations is not a novel idea [20,26,4] but the analysis we develop here seems to treat some details of this phenomenon in a complementary way.

*How does the buffering affect the sympathovagal balance?*

An interesting consequence of the buffering may be shown when the joint dynamics of the blood pressure and the heart rate is considered. The ABP is known to depend on the parameters of ventilation: i.e. the instantaneous lung volume (see e.g. [14,13]) and the current breathing phase, related with the respiratory frequency, inspiratory and expiratory time. The purpose of RSA seems to be the buffering of the respiratory related changes of the ABP in order to maintain a constant MAP. If this buffering is effective there will be activity in HF band of the HR. If it is ineffective, this activity will be visible in the HF band of ABP. During a pronounced adrenergic drive, HR is high and quite constant. In consequence, RSA disappears and the power in the HF band of the HR as well as the SDNN are greatly reduced. Therefore, in states of a pronounced adrenergic drive, as well as during the stimulation of the heart with a constant rate, the HF power of ABP must inevitably rise as the fluctuations are no longer buffered. Administration of beta-blockers that suppress the adrenergic drive will then shift the fluctuations power from the HF band of ABP towards the HF band of HR.

This introduces a kind of complex sympathovagal balance which is, in fact, governed by the sympathetic tone. This tone seems to decide if activity in the HF band will show up in ABP or in HR power spectrum. This balance seems to be strongly related to the buffering as a central phenomenon and affected by all the factors that influence the buffering, such as e.g. the position of the body. Note also that, if the assumption of the role of RSA is correct, the power in the HF band should be related with the stroke volume modulation depth and subject to all the already mentioned factors that may influence this modulation. The long discussions (e.g. [29] and references therein) that followed the formulation of the idea of sympathovagal balance seem to suggest that this balance is quite complex. As already shown e.g. in ref. [29] and [30], the HF and LF activity of the HR seem to be related in a complex way and, indeed, the mechanism that we propose seems to be complex enough to justify this observation. And indeed, the fact that the amplitude of RSA and the HF band activity of the BP depend strongly on sympathetic tone has been experimentally observed [4,30].

*Role of the baroreflex in the control of ABP fluctuations in the HF band*

Another interesting topic is the role of the baroreflex in the phenomenon of buffering which at first glance seems obvious: the HR controls the BP through the baroreflex. There exists, however, an experimental indication that the baroreflex is active only below the respiratory frequency: the chronic sinoaortic denervation does not affect the HF band of the ABP in rats [31].

Indeed, the efferent vagal activity seems rather to provide the information on the breathing phase and the tidal volume. The latter variable is known to strongly affect the depth of the respiratory modulation of the stroke volume [14]. Therefore, vagal activity would be merely a direct measure of the quantity which modulates the ABP. In physiological conditions, such a feedback would be sufficient for the estimation of the magnitude of the fluctuations themselves. One may thus hypothesize that in the process of buffering of the respiratory-related fluctuations of blood pressure, the information on the exact magnitude of the blood pressure fluctuations is actually not directly supplied to the effector. In fact, if it was supplied, the paradoxical enhancement of fluctuations obtained in our model and observed in ref. [2] would be rather impossible as the buffering system would rather minimize the amplitude of the control in order not to increase the fluctuations. In such cases when the HR paradoxically enhances the ABP fluctuations the vagal actual feedback appears as insufficient. The purely theoretical scope of this paper enables us only to speculate on this subject, but this speculations seem quite legitimate and concordant with the experimental results [2,31,13,14]. An interesting paper, pointing at the possibility of the involvement of the baroreceptor in the generation of RSA [11], in our opinion shows that the baroreflex has a potency of buffering the changes in the HF band. There are experiments [31] that suggest that, at least in rats, this does not occur in practice.

From the above hypothesis one may conclude that the ABP fluctuations are buffered by the HR fluctuations in two quite independent reflexes acting in different frequency ranges: the vagally mediated RSA in the HF band, and the sympathetically mediated baroreflex below the HF band. Accordingly, the feedback in the HF band would be provided by the vagal activity related to the respiratory frequency and tidal volume, while below the HF band it would be primarily due to the sympathetically mediated input from the baroreceptors. It also appears that the system that controls the ABP fluctuations in the HR band is tuned to work effectively in the upright position of the body. This limits the range of species, in which the described phenomenon may be observed, to those that maintain the upright position of the body.

The fact that the introduction of delay into the control loop may have a destabilizing effect was already studied theoretically [24,32,33]. The mechanism of instability, which appeared there, was a delay-dependent supercritical Hopf bifurcation which destabilizes a fixed point (or constant) solution and introduces oscillations with a frequency related to the delay length. The mechanism of the delay-induced instability that we present here is actually much simpler and does not introduce any additional frequency.

*Baroreflex sensitivity assessment: choice of variable*

From the evidence shown above, two methodological conclusions can be drawn.

Firstly, the analysis of the baroreflex loop may give different results depending on which variable derived from the ABP is considered: the mean or the systolic. We showed above that controlling ABP does not mean that all its dynamical variables are controlled to the same extent. Therefore it may be discussed which of the variables should be treated as the target of the control: which of them is actually measured by the baroreceptors and which of them is the most important for homeostasis to be maintained. Our study seems to yield a new understanding of the effect, well known from the experimental evidence of [2,3,4,34], that the dynamics of MAP and SAP is not always similar. Hence, MAP seems a better choice as the respiratory-related fluctuations of this variable are to a lesser extent dependent on the position of the body, which is especially important in the ambulatory blood pressure monitoring.

Secondly, an unexpected methodological conclusion is that the action of the baroreflex loop should be assessed in the upright position where the buffering is effective and not in the supine position. In the latter case, some of the blood pressure variability is actually introduced by heart rate variability itself which leads to an ineffective control of the ABP fluctuations. The dependence of the buffering on the position of the body was observed by the researchers who study directionality of the causal relation between the BP and the HR using the transfer function approach. They observed that the phase of the transfer function between the fluctuations in HR and in BP depends on the position of the body and that in the supine position the changes in SAP seem to be followed by the changes in HR [4,34]. This seems to be in a good agreement with our results, although the topic requires further experimental study.

In the above discussion, we considered the ABP fluctuations at respiratory frequency as a by-product of the respiration. In fact, the same argumentation may be applied to the low frequency (LF) range and to the ABP fluctuations that appear there. Also in this range, the activity of HR and ABP are known to be related [29,35], which may be interpreted as the effect of buffering. This opens the question whether the LF heart rate fluctuations may also have a destabilizing role under certain circumstances such as an increased circulation time, for example.

## Conclusions

A simple model of cardiovascular variability was proposed which shows a possible mechanism of the buffering of the respiratory-related changes of ABP. The model explains the selective buffering of ABP fluctuations obtained experimentally in [2] and the paradoxical enhancement of the ABP fluctuations, subject to changes of the position of the body observed in [4,34] as a result of an ineffective control. Several hypotheses were inferred from the results of this model. We discuss the balance between the HF fluctuations in ABP and in HR and postulate an important role of the tonic sympathetic activity in this phenomenon. We suggest that buffering of the BP fluctuations is a multilateral phenomenon: the buffering in the HF band seems to be provided by a mechanism, which acts through the vagal activity, distinct from the arterial baroreflex acting below the HF band. We formulate a hypothesis that information on the magnitude of the BP fluctuations in the HF range is not fed back directly to the heart. Due to this, in conditions such as the supine position of the body, the controlling mechanism additionally destabilizes the BP fluctuations. Finally, we discuss methodological issues: we propose to use the MAP as the proper variable for the assessment of the baroreflex sensitivity index and also propose to assess baroreflex sensitivity in the upright position of the body.

## Acknowledgement

The paper was supported by the Polish Ministry of Science, Grant No. 496/N-COST/2009/0.